\documentclass[paper=a4, 12pt]{article}

\usepackage{orcidlink}
\usepackage{polyglossia}
\setmainlanguage{english}
\usepackage{mathtools}
\usepackage{amsmath}
\usepackage{amssymb}
\usepackage{booktabs}
\usepackage{makecell}
\usepackage{todonotes}
\usepackage{biblatex}
\usepackage[default]{fontsetup}
\usepackage{microtype}
\usepackage{hyperref}
\usepackage[abbreviations]{glossaries-extra}

\bibliography{references}

\DeclareMathOperator{\PG}{PG}
\DeclareMathOperator{\CS}{CS}
\DeclareMathOperator{\TS}{TS}
\DeclareMathOperator{\AP}{AP}
\DeclareMathOperator{\Var}{Var}
\DeclareMathOperator{\Chan}{Chan}
\DeclareMathOperator{\Effect}{Effect}
\DeclareMathOperator{\Cond}{Cond}

\DeclareMathOperator{\Eval}{Eval}
\DeclareMathOperator{\Act}{Act}
\DeclareMathOperator{\Loc}{Loc}
\DeclareMathOperator{\Pre}{Pre}
\DeclareMathOperator{\Post}{Post}
\DeclareMathOperator{\Clocks}{Clocks}
\DeclareMathOperator{\dom}{dom}
\DeclareMathOperator{\id}{id}
\DeclareMathOperator{\qint}{int}
\DeclareMathOperator{\qext}{ext}
\DeclareMathOperator{\event}{ev}
\DeclareMathOperator{\origin}{orig}
\DeclareMathOperator{\Until}{U}

\newcommand{\longto}{\ensuremath{\longrightarrow}}
\newcommand{\TSdef}{\ensuremath{(S, \Act, \longto, I, \AP, L)}}
\newcommand{\PGdef}{\ensuremath{(\Loc, \Act, \Effect, \hookrightarrow, \Loc_0 , g_0)}}
\newcommand{\PTS}{\ensuremath{\mbox{\emph{PTS}}}}
\newcommand{\PTSdef}{\ensuremath{(S, \Act, p, I, \AP, L)}}
\newcommand{\TTS}{\ensuremath{\mbox{\emph{TTS}}}}
\newcommand{\TTSdef}{\ensuremath{(S, \Act, \longto, I, X, \AP, L)}}
\newcommand{\PTTS}{\ensuremath{\mbox{\emph{PTTS}}}}
\newcommand{\PTTSdef}{\ensuremath{(S, \Act, p, I, X, \AP, L)}}
\newcommand{\PPG}{\ensuremath{\mbox{\emph{PPG}}}}
\newcommand{\PPGdef}{\ensuremath{(\Loc, \Act, \Effect_p, \hookrightarrow, \Loc_0 , g_0)}}
\newcommand{\TPG}{\ensuremath{\mbox{\emph{TPG}}}}
\newcommand{\TPGdef}{\ensuremath{(\Loc, \Act, \Effect, \hookrightarrow, \Loc_0 , g_0)}}

\newcommand{\PTPGdef}{\ensuremath{(\Loc, \Act, \Effect_p, \hookrightarrow, \Loc_0 , g_0)}}

\newabbreviation{ts}{TS}{transition system}
\newabbreviation{pg}{PG}{program graph}
\newabbreviation{cs}{CS}{channel system}
\newabbreviation{tpts}{TPTS}{timed probabilistic transition system}
\newabbreviation{tppg}{TPPG}{timed probabilistic program graph}
\newabbreviation{tpcs}{TPCS}{timed probabilistic channel system}
\newabbreviation{sc}{SC}{state chart}
\newabbreviation{xml}{XML}{extensible markup language}
\newabbreviation{scxml}{SCXML}{state chart \glsps{xml}}
\newabbreviation{jani}{JANI}{JSON automata network interface}
\newabbreviation{scan}{SCAN}{statistical analyzer}

\makeindex

\title{The SCAN Statistical Model Checker}
\author{Enrico~Ghiorzi\thanks{Università di Genova \url{enrico.ghiorzi@edu.unige.it} \orcidlinkc{0000-0001-5983-6230}}
\and Armando~Tacchella\thanks{Università di Genova, \url{armando.tacchella@unige.it}\orcidlinkc{0000-0001-9487-331X}}}

\begin{document}
\maketitle

\begin{abstract}
	This paper lays out the formal foundations upon which the \glsfmtshort{scan} statistical model checker is built.
	% \keywords{Verification \and Statistical model checking \and Channel system}
\end{abstract}

\section{Background}%
\label{sec:background}

\begin{figure}
	\centering
	\scalebox{0.4}{\includegraphics{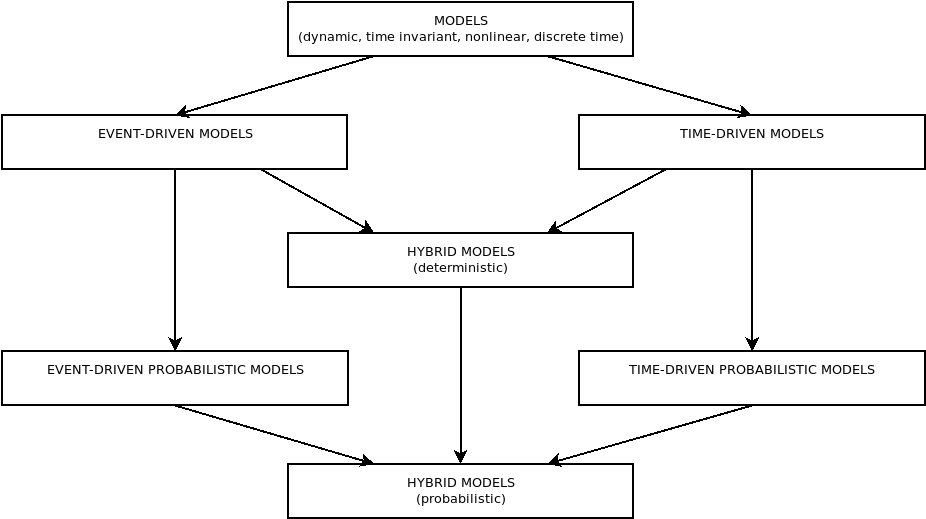}}
	\caption{\label{fig:model_taxonomy} A taxonomy of models.}
\end{figure}

Here we briefly review the major model classifications that are
relevant throughout this document. The system we model is a robot
together with its control software at various layers and its hardware
interacting with an external environment. The scenario is
intrinsically \emph{dynamic} and we postulate that it is
also \emph{time invariant}, i.e., the parameters governing
the overall system behavior do not change significantly over
time and random processes are characterized by stationary
distributions. We expect that \emph{nonlinear} models will have to be
considered in order to take into account all kinds of interactions between
the robot and the environment, but we restrict ourselves to
\emph{discrete time} models, i.e., in all cases where time needs to
be considered explicitly (see below), we assume that it does so in
discrete --- albeit possibly very small --- steps. The motivation for
this choice is that, from the perspective of control software design,
most of the signals travelling among the various components are either
intrinsically discrete or made available at specific sampling rates.
The implication is that we will model the physical world using
\emph{difference equations} rather than \emph{differential
	equations}. Since our models will have to be executed, this spares us
many potential issues when it comes to determine the trajectories of state
variables subject to such equations.
Other elements for which we may consider alternatives are:
\begin{itemize}
	\item \textbf{Continuous-state vs.\ Discrete-state}: we expect that
	      a combination of continuous and discrete quantities might
	      characterize the overall state of the system; whenever possible,
	      we will prefer discrete and finite domains for variables in order
	      to preserve computational tractability of, e.g., algorithmic model
	      verification.
	\item \textbf{Time-driven vs.\ Event-driven}: in time-driven systems,
	      the state changes as time changes. The models of physical
	      interactions, e.g., between the robot and the environment are
	      typically modeled in a time-driven fashion. In event-driven
	      systems, it is only the occurrence of asynchronously generated
	      discrete events that forces instantaneous state transitions. The
	      models of software interactions, e.g., among different control
	      software components, are usually best modeled as such. Since we
	      deal with combinations of software, hardware and environment we
	      expect that \emph{hybrid models}, featuring both time-driven and
	      event-driven models, will have to be considered.
	\item \textbf{Deterministic vs.\ Stochastic evolution}: a model is
	      stochastic whenever one or more of its parameters is a random
	      variable. In this case, the state of the system is described by a
	      stochastic process, and a probabilistic framework is required to
	      characterize the model behavior.
\end{itemize}
Leaving the distinction between continuous-state and finite-state
aside, in Figure~\ref{fig:model_taxonomy} we describe the taxonomy of the
models we consider.

\section{\texorpdfstring{\Glsfmtlongpl{tpts}}{Timed Probabilistic Transition Systems}}

A \emph{(labelled) transition system} \(\TS\) is a tuple
\(\TSdef\) where:
\begin{itemize}
	\item $S$ is a set of \emph{states};
	\item $\Act$ is a set of \emph{actions};
	\item $\longto \subseteq S \times \Act \times S$ is a \emph{transition
		      relation};
	\item $I \subseteq S$ is a set of \emph{initial states};
	\item $\AP$ is a set of \emph{atomic propositions}, and
	\item $L : S \to 2^{\AP}$ is a \emph{labeling function}.\footnote{Given
		      a set $A$, we denote with $2^A$ the \emph{powerset} of $A$, i.e.,
		      the set of all subsets of $A$.}
\end{itemize}
The labeling function $L$ relates a set $L(s) \in 2^{\AP}$ of atomic
propositions to any state $s$. $L(s)$ stands for exactly those atomic
propositions $a \in \AP$ which are satisfied by state $s$. Given
a propositional logic formula $\varphi$, $s$ \emph{satisfies} the
formula $\varphi$ if the evaluation induced by $L(s)$ makes the
formula $\varphi$ true. We write $s \models \varphi$ when $L(s)
	\models \varphi$.
A \(\TS\) is \emph{finite} if $S$, $\Act$ and $\AP$ are finite.
We call \emph{transitions} the triples in the transition relation and
we write \(s_0 \xrightarrow{\alpha} s_1\) in place of \((s_0, \alpha, s_1) \in
\longto\); in a transition $s_0 \xrightarrow{\alpha} s_1$, $s_0$ is
the \emph{pre-state} (or \emph{source state}) and $s_1$ is the
\emph{post-state} (or \emph{target state}).
The set of all post-states from a state $s$ is defined as
\[
	\Post(s,\alpha) = \{ s' \in S \mid s \xrightarrow{\alpha} s' \}
	\qquad
	\Post(s) = \bigcup_{a \in \Act} \Post(s, \alpha)
\]
Analogously, the set of all pre-states of $s$ is defined as
\[
	\Pre(s,\alpha) = \{ s' \in S \mid s' \xrightarrow{\alpha} s \}
	\qquad
	\Pre(s) = \bigcup_{a \in \Act} \Pre(s, \alpha)
\]
The above notation can be extended to sets of states in the obvious
way. A state  $s \in S$ is \emph{terminal} if $Post(s) = \emptyset$. A
transitions system is \emph{action-deterministic} if $|I| \leq 1$ and
$|Post(s,\alpha)| \leq 1$ for all $s \in S$ and $\alpha \in \Act$; it is
\emph{AP-deterministic} if we have that
\(\lvert\{s \in I \mid L(s) = A\}\rvert \leq 1\) and
\[
	\lvert\Post(s,\alpha) \cap \{s' \in S \mid L(s') = A\}\rvert \leq 1
\]
for all \(A \in 2^{\AP}\).

Let $\TS = \TSdef$ be a transition system. A \emph{(finite) execution
	fragment} $\rho$ of \(\TS\) is an alternating sequence of states and
actions ending with a state
\[
	\rho = s_0 \alpha_1 s_1 \alpha_2 \ldots \alpha_n s_n \mbox{ such that }
	s_i \xrightarrow{\alpha_{i+1}} s_{i+1} \mbox{ for all } 0 \leq i < n
\]
where $n$ is the \emph{length} of the execution fragment.
The sequence with a single $s \in S$ is an execution fragment
of length $n=0$ according to the definition above. An execution
fragment is \emph{infinite} when it has infinite length. Execution
fragments will be written as
\[
	\rho = s_0 \xrightarrow{\alpha_1} s_1 \xrightarrow{\alpha_2} \ldots
	\xrightarrow{\alpha_n} s_n \qquad \sigma = s_0 \xrightarrow{\alpha_1}
	s_1 \xrightarrow{\alpha_2} \ldots
\]
where $\rho$ is finite and $\sigma$ is infinite. An execution fragment
is \emph{maximal} if it is either infinite or it ends in a terminal
state. An execution fragment is \emph{initial} if it starts in an
initial state. An \emph{execution} of a \(\TS\) is an initial, maximal
execution fragment. A state \(s\) is said to be \emph{reachable} if
there is an execution fragment that ends in $s$ and that starts in
some initial state $i \in I$:
\[
	i \xrightarrow{\alpha_1} s_1 \xrightarrow{\alpha_2} \ldots
	\xrightarrow{\alpha_n} s_n = s
\]
We let $Reach(\TS)$ denote the set of reachable
states in \(\TS\). Finally, a (deterministic) \emph{policy} (or
\emph{scheduler}) is any mapping $\pi_{\TS} : S \to \Act$. Intuitively,
a policy decides which action is performed in each state. When the
\(\TS\) to which we refer is clear from the context we simply write
$\pi$ to denote a policy.

A \emph{(labelled) probabilistic transition system} \PTS{} is a tuple \PTSdef{}
where $S$, $\Act$, $I$, $AP$ and $L$ are defined as in standard
transitions systems and $p$ is the \emph{transition probability} $p: S
	\times \Act \times S \to [0,1]$. Informally, $p(s, \alpha, s')$ defines
the probability that state $s'$ is reached from state $s$ when the
action $\alpha$ is performed. Transition probabilities are subject to
the following sanity condition:
\[
	\sum_{s' \in S} p(s,\alpha,s') = 1 \qquad \mbox{for all } s \in S,
	\alpha \in \Act
\]
Notice that, modulo the definition of a \emph{reward function}, the
notion of \PTS{} corresponds to that of a (stationary) Markov decision
process (MDP). The terminology that we introduced for transition
systems carries over to their probabilistic variant, with some
required adaptations:
\begin{itemize}
	\item A transition $s \xrightarrow{\alpha} s'$ exists whenever
	      $p(s,\alpha,s') > 0$, where $s$ is the pre-state and $s'$ is the
	      post-state.
	\item Given a specific action $\alpha$, the set of all post-states
	      from a state $s$ is $\Post(s,\alpha) = \{s' \in S \mid s \xrightarrow{\alpha} s' \}$; if all actions are considered, the set of all post-states
	      is defined as $\Post(s) = \bigcup_{\alpha \in \Act} \Post(s, \alpha)$;
	      similar adaptations are required to define pre-states.
	\item A state $s \in S$ is terminal when $\Post(s) = \emptyset$.
	\item An execution fragment is an alternating sequence of
	      states and actions ending with a state in case it is finite:
	      \[
		      \rho = s_0 \xrightarrow{\alpha_1} s_1 \xrightarrow{\alpha_2}
		      \ldots \xrightarrow{\alpha_n} s_n \qquad \sigma = s_0 \xrightarrow{\alpha_1}s_1 \xrightarrow{\alpha_2} \ldots
	      \]
	      Notice that the existence of the transitions $s_i \xrightarrow{\alpha_{i+1}} s_{i+1}$ implies that $p(s_i,\alpha,s_{i+1}) > 0$ for all $i > 0$.
	\item Initial and maximal execution fragments are defined as in
	      \(\TS\); an execution of a \PTS{} is an initial maximal execution
	      fragment.
\end{itemize}
The stochastic process generated by the interaction of a probabilistic
transition system $\PTSdef$ and a policy $\pi$ can be formalized as
follows. Let us define the \emph{sample space} $\Omega$ as
\[
	\Omega = S \times A \times S \times A \times S \times A
	\times \cdots = {\{ S \times A \}}^{\infty}
\]
where members of $\Omega$ are infinite execution fragments.
If we let $P_1(\cdot)$ be some \textit{initial distribution} of the
system state, the \emph{probability of an execution fragment} $\rho =
	(s_1, a_1, s_2, a_2, \ldots, s_k)$, $k \leq \infty$,  induced by a
policy $\pi$ can be defined as
\begin{equation}
	P^\pi(\rho) =
	P_1(s_1)p(s_1,\pi(s_1),s_2)p(s_2, \pi(s_2), s_3) \ldots
	p(s_{k-1}, \pi(s_{k-1}), s_k)
\end{equation}
The triple $\{\Omega, B(\Omega), P^\pi\}$ is the \textit{probability
	model} defined by the policy $\pi$ in the probabilistic transition
system, where $B(\Omega)$ denotes a $\sigma$-algebra of (Borel)
measurable subsets of $\Omega$, and $P^\pi$ is indeed a probability
measure on $B(\Omega)$. Given such model, we can define two random variables
$X_t$ and $Y_t$ to take values in $S$ and $A$, respectively, by
\begin{equation}
	X_i(\rho) = s_i \quad \mbox{and} \quad Y_i(\rho) = a_i = \pi(s_i)
\end{equation}
for $i = 1, 2, \ldots, k$, and $k \leq \infty$.
When the execution fragment is
$\rho$, the random variable $X_i$ denotes the $i$-th state, and
$Y_i$ denotes the $i$-th action chosen (deterministically) by
$\pi$. Given a \PTS{} and a policy $\pi$
the property
\begin{equation}
	\label{eq:markovchain}
	P^\pi(a_i, s_{i+1}, \ldots, s_k | s_1, a_1, \ldots, s_i) = P^\pi(a_i,
	s_{i+1}, \ldots s_k | s_i)
\end{equation}
is true independently of the specific actions decided by
$\pi$, i.e., the stochastic process $\{ X_i; i \in \mathbb{Q^+} \}$
induced by a policy $\pi$ on a \PTS{} is a \emph{discrete time Markov chain} --- see~\cite{puterman2009markov}.

A \emph{(labelled) timed transition system} \TTS{} is defined as a a tuple
\TTSdef{} where $S$, $\Act$, $I$, $\AP$ and $L$ are defined as previously
and
\begin{itemize}
	\item $X$ is a finite set of \emph{clock variables} (also
	      \emph{clocks}) that take values over the domain $\mathbb{Q}^+$ of
	      positive rational numbers (including $0$);
	\item $\longto \subseteq S \times \Act \times 2^X \times \Phi(X) \times
		      S$ is a \emph{timed transition relation}; $\Phi(X)$ denotes the set of
	      allowable \emph{clock constraints} on the clock variables $X$ and
	      it is limited to expressions of the form
	      \[
		      \Phi(X) = x \leq c \mid c \leq x \mid \neg \phi_1 \mid \phi_1
		      \wedge \phi_2
	      \]
	      where $x \in X$, $c \in \mathbb{Q}^+$ and $\phi_1, \phi_2 \in
		      \Phi(X)$; additional constraints can be defined as abbreviations:
	      \begin{itemize}
		      \item $\phi_1 \vee \phi_2 := \neg (\neg \phi_1 \wedge \neg \phi_2)$;
		      \item $\mathbf{true} := \phi_1 \vee \neg \phi_1$;
		      \item $\mathbf{false} := \phi_1 \wedge \neg \phi_1$;
		      \item $x = c := x \leq c \wedge c \leq x$
	      \end{itemize}
	      where $\phi_1, \phi_2$ are arbitrary constraints, $x \in X$ and
	      $c \in \mathbb{Q}^+$.
\end{itemize}
Notice that our definition of \TTS{} implies a discrete time model
since the clocks are defined over the domain of rationals which is not
dense, i.e., $\mathbb{Q} \subset \mathbb{R}$.
Unless explicitly defined otherwise in the following, the terminology
for transition systems carries over to timed transition systems.
We call \emph{timed transitions} the tuples in ``$\longto$'' and,
assuming $Y \subseteq X$ and $\phi \in \Phi(X)$,
we write $s_0 \xrightarrow{\alpha, Y, \phi} s_1$ in place of
$(s_0, \alpha, Y, \phi, s_1) \in \longto$; in a transition $s_0
	\xrightarrow{\alpha, Y, \phi} s_1$, $s_0$ is the pre-state
and $s_1$ is the post-state; $Y$ is the subset of clock variables which is
\emph{reset} to $0$ when the transition occurs and $\phi$ is the clock
constraint \emph{guarding} the transition, i.e., the transition will not
occur unless $\phi$ is satisfied given the current values of the clock
variables occurring in $\phi$. At any instant, the reading of a clock
equals the time elapsed since the last time it was reset and all
clocks proceed at the same rate. Intuitively, clocks can be
seen as stop-watches which can be started and checked independently of
one another, but all the clocks increase at the uniform rate counting
time with respect to a global (discrete) time
frame~\cite{DBLP:journals/tcs/AlurD94}.
A \emph{clock interpretation} $\nu$ for a set $X$ of clocks assigns a
value to each clock, i.e., $\nu : X \to \mathbb{Q}^+$. Given a clock
interpretation $\nu$ and a constraint $\phi$, we write $\nu
	\models \phi$ to denote that $\nu$ satisfies $\phi$.
Given a clock interpretation $\nu$, the set of all post-states from a
state $s$ is defined as
\[
	\Post(s,\alpha,\nu) = \{ s' \in S \mid s \xrightarrow{\alpha,Y,\phi} s'
	\mid \nu \models \phi \}
	\qquad
	\Post(s,\nu) = \bigcup_{a \in \Act} \Post(s, \alpha, \nu)
\]
Analogously, the set of all pre-states of $s$ is defined as
\[
	\Pre(s,\alpha,\nu) = \{ s' \in S \mid s' \xrightarrow{\alpha,Y,\phi} s
	\mid \nu \models \phi\}
	\qquad
	\Pre(s,\nu) = \bigcup_{a \in \Act} \Pre(s, \alpha, \nu)
\]
The above notation can be extended to sets of states in the obvious
way. A state  $s \in S$ is \emph{terminal} if $\Post(s, \nu) =
	\emptyset$ for all possible clock interpretations $\nu$. A
transitions system is \emph{action-deterministic} if $|I| \leq 1$ and
$|\Post(s,\alpha,\nu)| \leq 1$ for all $s \in S$, $\alpha \in Act$ and
clock interpretations $\nu$; it is
\(\AP\)-\emph{deterministic} if $|I| \leq 1$ and
\[
	|\Post(s,\alpha,\nu) \cap \{s' \in S \mid L(s') = a\}| \leq 1
\]
for all atomic propositions $a \in \AP$ and for all possible clock
interpretations $\nu$.

Let $\TTS = \TTSdef$ be a timed transition system. For $t \in
	\mathbb{Q}^+$, $\nu + t$ denotes the clock interpretation which maps
every clock to the value $\nu(x) + t$.
For $Y \subseteq X$, we denote with $[Y \mapsto t]\nu$ the clock
interpretation for $X$ which assigns $t$ to each $x \in Y$ and agrees
with $\nu$ over the rest of the clocks. Finally, let $\tau = \tau_1
	\tau_2, \ldots$ be a \emph{time sequence}, i.e., an infinite sequence
of time values $\tau_i \in \mathbb{Q}^+$ with $\tau_i > 0$ such that:
\begin{itemize}
	\item $\tau$ increases monotonically, i.e., $\tau_i \leq \tau_{i+1}$
	      for all $i \geq 1$; this condition ensures that the logical
	      order of the time values is consistent with the temporal order,
	      while permitting adjacent observations with the same time stamp.
	\item $\tau$ progresses, i.e., for every time $t \in \mathbb{Q}^+$
	      there is some $i \geq 1$ such that $\tau_i > t$.
\end{itemize}
An \emph{execution} $\sigma$ of a \TTS{} is a sequence of the form
\[
	\sigma = \langle s_0, \nu_0 \rangle \xrightarrow{\alpha_1, \tau_1}
	\langle s_1, \nu_1 \rangle \xrightarrow{\alpha_2, \tau_2}
	\langle s_2, \nu_2 \rangle \xrightarrow{\alpha_3, \tau_3}\ldots
\]
with $s_i \in S$ and $\nu_i : X \to \mathbb{Q}^+$ for all $i \geq 0$,
satisfying the following requirements:
\begin{itemize}
	\item \emph{Initiation}: $s_0 \in I$ and $\nu_0(x) = 0$ for all $x
		      \in X$.
	\item \emph{Consecution}: for all $i \geq 1$, there is a transition
	      of the form $s_{i-1} \xrightarrow{\alpha, Y, \phi} s_i$ such that
	      $\nu_{i-1} + (\tau_i - \tau_{i-1}) \models \phi$ and
	      $\nu_i = [Y \mapsto 0](\nu_{i-1} + (\tau_i - \tau_{i-1}))$.
\end{itemize}
Intuitively, the execution of a timed transition system is a
sequence of actions (events) occurring at specific instants in time
(timestamps) that are compatible with the clock guards. The choice of
rationals for time values allows us to consider both integer and
fractional time stamps and thus handle different time scales with
reference to the global time frame. The \emph{finite projection}
$\rho$ of an execution $\sigma$ of a \TTS{} is the finite sequence
\[
	\rho = \sigma_n = \langle s_0, \nu_0 \rangle \xrightarrow{\alpha_1, \tau_1}
	\langle s_1, \nu_1 \rangle \xrightarrow{\alpha_2, \tau_2} \ldots
	\xrightarrow{\alpha_{n-1}, \tau_{n-1}} \langle s_{n-1}, \nu_{n-1}
	\rangle \xrightarrow{\alpha_n, \tau_n}
	\langle s_n, \nu_n \rangle
\]
which contains only the first $n+1$ pairs of state and clock
evaluations contained in $\sigma$.
A state $s \in S$ is said to be \emph{reachable} if
there is a finite projection $\rho = \sigma_n$ for some execution
$\sigma$ such that contains $s = s_n$. We let $Reach(\TTS)$
denote the set of reachable states in \TTS.

The combination of a \PTS{} and a \TTS{} yields a \emph{probabilistic
	timed transition system} (\PTTS{}). The features of such a model are
best understood in terms of a \TTS{} extended with
probabilities. Intuitively, while a timed transition system defines
exactly the state(s) that an action may lead to given the current
state, a probabilistic timed transitions system defines the successor
state(s) in term of probabilities. Formally, a \PTTS{} is a tuple
\PTTSdef{} where all the elements have the same meaning as in the
definition of a \TTS{} and $p$ is a \emph{transition probability}
defined as
\[
	p : S \times \Act \times 2^X \times \Phi(X) \times S \to [0,1].
\]
Informally, $p(s,\alpha, Y, \phi, s')$ defines the probability that
state $s'$ is reached from a state $s$ when the action $\alpha$
subject to the clock constraint $\phi$ and implying the reset of the
clocks in $Y$ is performed. Given any triplet $(\alpha, Y, \phi) \in
	\Act \times 2^X \times \Phi(X)$ we impose the additional sanity condition on $p$
that:
\[
	\sum_{s' \in S} p(s,\alpha,Y,\phi,s') = 1 \mbox{ for all } s \in S
\]
In other words, every action, together with its associated clock resets
and conditions, determines a distribution of successor states;
different actions, even from the same state, may have different (and
unrelated) distributions of successor states.
A transition, denoted as
$s_0 \xrightarrow{\alpha, Y, \phi} s_1$ is defined when $p(s,\alpha, Y,
	\phi,s') > 0$.  Given this notation, the execution of a \PTTS{} is
defined exactly as an execution of a \TTS{}. The other
definitions carry over from timed transition systems,
once the notion of transition is updated as above.

\begin{table}
	\footnotesize
	% \begin{center}
	\begin{tabular}{llll}
		\toprule
		Model & Definition & Dynamic (informal)       & Dynamic (formal)                                                  \\ \midrule
		TS    & \TSdef     & Transition Relation      & $\longto \subseteq S \times \Act \times S$
		\\ \midrule
		PTS   & \PTSdef    & Transition Probability   & $p: S \times \Act \times S \to [0,1]$
		\\ \midrule
		TTS   & \TTSdef    & Timed Trans. Relation    & $\longto \subseteq S \times \Act \times 2^X \times \Phi(X) \times
			S$
		\\ \midrule
		PTTS  & \PTTSdef   & Timed Trans. Probability & $p : S \times \Act \times 2^X \times \Phi(X) \times S \to [0,1]$
		\\ \midrule \midrule
		\multicolumn{4}{l}{$S$: set of states --- $\Act$: set of actions
			--- $I \subseteq S$: initial states}
		\\
		\multicolumn{4}{l}{$AP$ atomic propositions --- $L : S \to 2^{\AP}$
			labeling function}
		\\
		\multicolumn{4}{l}{$X$: set of clock variables --- $\Phi(X)$: set of clock constraints}
		\\ \bottomrule
	\end{tabular}
	% \end{center}
	\caption{\label{tab:ts_synopsis} Synopsis of transitions systems.}
\end{table}

The exact description of the stochastic process generated by the
interaction of a probabilistic timed transition system and a policy is
beyond the scope of this introductory text. However, it can be shown
that applying a scheduler to a \PTTS{} yields a kind of Markov chain
wherein the transitions occur at specific instants in time determined
by the clock constraints and the progress of the clocks in each
state. Such a process can be described within the framework of
\emph{piecewise deterministic Markov processes} (PDMP), a kind of
hybrid system whose state evolves according to ordinary differential
equations at specified locations, but jumps among locations may occur
at given instants in time; both such instants and the target locations
are drawn from random distributions in the general PDMP framework. In
particular, the dynamics of a PDMP are defined through specifying
three quantities~\cite{Fearnhead_2018}:
\begin{itemize}
	\item the \emph{deterministic dynamics} that characterize the
	      system at any given location;
	\item the \emph{event rate} that characterizes the time distribution
	      in which events (change of location) occur;
	\item the \emph{transition distribution at events} that
	      characterizes  the location distribution when an event occurs.
\end{itemize}
In our case, the deterministic dynamics correspond to the evolution of
the clock variabiles at a constant rate, the event rate is
deterministic and can be computed considering the policy and the clock
constraints (an action can be taken in a state when the clock
constraints are satisfied), and the transition distribution is given
by computing from the transition probability of the \PTTS{} the
probability of reaching a state given the current one and the action
suggested by the policy.

Table~\ref{tab:ts_synopsis} summarizes all the different kinds of
transition systems that we have introduced. For each model, we report
the acronym and the definition of the relation that governs the
dynamics of the model.

\section{Timed Probabilistic Program Graphs and Timed Probabilistic Channel Systems}%
\label{sec:tpcs}

\subsection{Program Graphs}%
\label{sec:program_graphs}

Transition systems are well suited to represent reactive computations,
but their formalism is focused on modeling the control structure of
systems. The executions of a data-dependent system typically result
from some kind of conditional branching based on values taken by
internal variables. In principle, this could be modeled with transition
systems by handling branching on specific values with nondetermism.
However, the resulting
model would be very abstract and only a few relevant properties could
be verified, if any~\cite{baier2008principles}.
Program graphs extend the formalism of transition
systems to express data-dependency of executions, but their semantics
can still be given in terms of transitions systems. Intuitively,
program graphs are the ``high level programming language'' useful for
modeling, whereas transition systems are the ``assembly language''
required to give precise semantics and enable verification. In
practice, the ``compilation'' of program graphs into transition
systems provides an operational semantics for program graphs.

Let $\Var$ be a set of typed variables. For each $x \in \Var$, let
$\dom(x)$ represent the \emph{domain} of the variable $x$, i.e., the
set of values that the variable can take, and let $\eta(x) \in \dom(x)$
represent the \emph{valuation} of $x \in \Var$.
Further, let $\Cond(\Var)$ be the set of Boolean conditions
over $\Var$, i.e., propositional logic formulas whose propositional
symbols are of the form ``$ \overline{x} \in \overline{D} $''
where $  \overline{x} = (x_1, \dots , x_n ) $ is a
tuple consisting of pairwise distinct variables in $\Var$ and
$D$ is a subset of $\dom(x_1) \times  \cdots \times \dom(x_n)$.
If $\Eval(\Var)$ is the set of (variable) evaluations that assign
values to variables, and $\Act$ is a set of actions, then the
\emph{effect function} $\Effect : \Eval(\Var) \times \Act \to
	\Eval(\Var)$ dictates how the evaluation of a variable changes because
an action is taken. A \emph{program graph} over a set $\Var$ of
typed variables is a digraph whose edges are labeled with conditions
on these variables and it is defined as
\[
	\PG = \PGdef
\]
where
\begin{itemize}
	\item $\Loc$ is a set of locations and $\Act$ is a set of actions,
	\item $\Effect : \Eval(\Var) \times \Act \rightarrow \Eval(\Var) $ is
	      the \emph{effect function},
	\item  $ \hookrightarrow \; \subseteq \; \Loc \times \Act \times
		      \Cond(\Var) \times \Loc $ is the  \emph{conditional transition relation},
	\item  $\Loc_0\subseteq Loc$ is a set of initial locations,
	\item  $g_0 \in \Cond(\Var)$ is the initial condition.
\end{itemize}
The notation $l \xhookrightarrow{g:\alpha} l'$ is used as a
shorthand for $(l, \alpha, g, l') \in \hookrightarrow$. The condition
$g$ is also called the \emph{guard} of the conditional transition $l
	\xhookrightarrow{g:\alpha} l'$. If the guard is a tautology, then we
write $l \xhookrightarrow{\alpha} l'$.

A \emph{probabilistic program graph (PPG)} is obtained by extending a standard
program graph with the ability to change the value of the variables
according to some probability distribution. The definition of a PPG is
the same as a standard PG, but the effect function is now
probabilistic. Formally, given a generic $x \in \Var$, we let $\Eval(x)$
be the set of possible evaluations of $x$ and we define a
\emph{probabilistic effect function}:
\[
	\Effect_p : \Eval(x) \times \Act \times \Eval(x) \to [0,1]
\]
subject to the sanity condition that, for all evaluations $\eta(x) \in
	\Eval(x)$ and for all actions $\alpha \in \Act$ we have
\[
	\sum_{\eta'(x) \in \Eval(x)} \Effect_p(\eta(x), \alpha, \eta'(x)) = 1
\]
A standard PG is simply a PPG where, for each action $\alpha \in \Act$,
given $\eta'(x) = \Effect(\eta(x), \alpha)$ we have that
$\Effect_p(\eta(x), \alpha, \eta'(x)) = 1$ and $\Effect_p(\eta(x),
	\alpha, \eta''(x)) = 0$ for all $\eta''(x) \in \Eval(x)$ such that
$\eta''(x) \neq \eta'(x)$.

\begin{table}
	\scriptsize
	% \begin{center}
	\begin{tabular}{lll}
		\toprule
		Model & Definition & Dynamic                                                                                                         \\ \midrule
		PG    & \PGdef     & $\Effect: \Eval(\Var) \times \Act \to \Eval(\Var)$
		\\
		      &            & $ \hookrightarrow \; \subseteq \; \Loc \times \Act \times
			\Cond(\Var) \times \Loc $
		\\ \midrule
		PPG   & \PPGdef    & $\Effect_p: \Eval(\Var) \times \Act \times \Eval(\Var) \to [0,1]$
		\\
		      &            & $ \hookrightarrow \; \subseteq \; \Loc \times \Act \times
			\Cond(\Var) \times \Loc $
		\\ \midrule
		TPG   & \TPGdef    & $\Eval_t : \Clocks \to \mathbb{Q}^+$
		\\
		      &            & $\Effect: \Eval(\Var) \times \Eval_t(\Clocks) \times \Act \to \Eval(\Var) \times \Eval_t(\Clocks)$
		\\
		      &            & $ \hookrightarrow \; \subseteq \; \Loc \times \Act \times
			\Cond(\Var) \times \Cond(\Clocks)\times \Loc $
		\\ \midrule
		PTPG  & \PTPGdef   & $\Eval_t : \Clocks \to \mathbb{Q}^+$
		\\
		      &            & $\Effect: \Eval(\Var) \times \Eval_t(\Clocks) \times \Act \times \Eval(\Var) \times \Eval_t(\Clocks) \to [0,1]$
		\\
		      &            & $ \hookrightarrow \; \subseteq \; \Loc \times \Act \times
			\Cond(\Var) \times \Cond(\Clocks)\times \Loc $
		\\ \midrule
		\multicolumn{3}{l}{$\Var$: set of variables --- $\Eval(Var)$: variable evaluations --- $\Cond(\Var)$: Boolean conditions over $\Var$}
		\\
		\multicolumn{3}{l}{$\Loc$: set of locations --- $\Act$: set of actions
			--- $\Loc_0 \subseteq \Loc $: initial locations}
		\\
		\multicolumn{3}{l}{$g_0 \in \Cond(\Var)$: initial condition --- $\Act$: set of actions
			--- $\Loc_0 \subseteq \Loc $: initial locations}
		\\
		\multicolumn{3}{l}{$\Clocks$: set of clock variables --- $\Cond(\Clocks)$: clock constraints over $\Clocks$}
		\\ \bottomrule
	\end{tabular}
	% \end{center}
	\caption{\label{tab:pg_synopsis} Synopsis of program graphs.}
\end{table}

A \emph{timed program graph (TPG)} is obtained by adding a set
of clock variables $\Clocks$ and a specific evaluation
function $\Eval_t: \Clocks \to \mathbb{Q}^+$. As in a TTS,
clock variables are all incremented at the same uniform rate, starting
from the value $0$ in the initial state, and they can be reset to zero
during transitions. The definition of the
effect function and the transition relation are modified as follows:
\begin{itemize}
	\item the effect function takes into account the current clock
	      evaluations and possibly resets some clock values to zero, i.e.,
	      the function $\Effect : \Eval(\Var) \times \Eval_t(\Clocks) \times \Act
		      \rightarrow  \Eval(\Var) \times \Eval_t(\Clocks)$ is subject to the
	      condition that, for every $\alpha \in \Act$, $\eta,\eta' \in
		      \Eval(\Var)$ and $\delta, \delta' \in \Eval_t(Clocks)$ such that
	      $(\eta',\delta') = \Effect(\eta, \delta, \alpha)$ we have
	      either $\delta'(x) = \delta(x)$ or $\delta(x') = 0$ for all $x \in \Clocks$;
	\item the conditional transition relation takes into account clock
	      guards as well, i.e., $\hookrightarrow \; \subseteq \; \Loc
		      \times \Act \times \Cond(\Var) \times \Cond_t(\Clocks)
		      \times \Loc$, where $\Cond_t(\Clocks)$ is the set of allowable
	      clock constraints as introduced in~\ref{sec:background}.
\end{itemize}

Finally, a \emph{probabilistic timed program graph (PTPG)} is a TPG
where the $\Effect$ function is also probabilistic. Informally, a PTPG
behaves like a TPG when it comes to time, i.e., $(i)$ transitions are
guarded by clock constraints and $(ii)$ clock variables can be reset across
transitions, but the effect function is probabilistic in the sense
that standard variables in $\Var$ change their value according to some
probability distribution. Formally,
\[
	\Effect_p : \Eval(Var) \times \Eval_t(\Clocks)
	\times \Act \times \Eval(Var) \times \Eval_t(\Clocks) \to [0,1]
\]
subject to the sanity condition that, given a generic $x \in \Var$, for all
evaluations $\eta(x) \in \Eval(x)$ and for all actions $\alpha \in
	\Act$ we have
\[
	\sum_{\eta'(x) \in \Eval(x)} \Effect_p(\eta(x), \cdot, \alpha,
	\eta'(x), \cdot) = 1
\]
where the ``$\cdot$'' in place of the clock evaluation means that this
happens for all pairs of current and next clock evaluations.

\subsection{Parallel Composition of Program Graphs (Channel Systems)}%
\label{sec:channel_systems}

A \emph{channel system} $ \CS $ over $ (\Var, \Chan) $ consists of program graphs
$ \PG_i $ over $ (\Var_i , \Chan) $
(for $ 1 \leqslant i \leqslant n $) with $ \Var = \bigcup_{1 \leqslant i \leqslant n} \Var_i $. We denote
\[
	\CS = [\PG_1 | \dots | \PG_n] .
\]

The transition relation $ \hookrightarrow $ of a program graph over $ (\Var, \Chan) $
consists of two types
of conditional transitions. The conditional transitions
\( \ell  \xhookrightarrow{g:\alpha} \:  \ell' \)
are labeled with guards and actions (the condition $ g $ is
the guard of the conditional transition $ g\colon\alpha $).
These conditional transitions can happen whenever the guard holds.
Alternatively, conditional transitions may be labeled with communication actions. This
yields conditional transitions of type
\( \ell  \xhookrightarrow{g:!(p,q,x)} \:  \ell' \)
(for sending a message $ x $ along channel $ (p,q) $),
\( \ell  \xhookrightarrow{g:!(p,q,{<}m{>})} \:  \ell' \)
(for sending the specific value $ m $ along channel $ (p,q) $),
\( \ell  \xhookrightarrow{g:?(q,p,x)} \:  \ell' \)
(for receiving a message along channel $ (p,q) $) and
\( \ell  \xhookrightarrow{g:?(q,p,{<}m{>})} \:  \ell' \)
(for receiving the specific value $ m $ along channel $ (p,q) $).
Based on the current variable evaluation, the capacity and the content
of the channel $ (p,q) $, we assume in the following that these
conditional transitions are executable, i.e., the guard is satisfied:
\begin{itemize}
	\item \emph{Handshaking}. If $ cap((p,q)) = 0 $, then process $ p $ can transmit
	      over channel $ (p,q) $
	      \begin{itemize}
		      \item a message $ x $ by performing: \hspace{4.5em}
		            $ \ell_p  \xhookrightarrow{!(p,q,x)} \:  \ell'_p $ \quad or
		      \item a specific value $ m $ by performing: \hspace{1.3em}
		            $ \ell_p  \xhookrightarrow{!(p,q,{<}m{>})} \:  \ell'_p $
	      \end{itemize}
	      only if another process $ q $ offers a complementary receive action, i.e., can perform
	      \begin{align*}
		       & \ell_q  \xhookrightarrow{?(q,p,x)} \:  \ell'_q \quad   or \\
		       & \ell_q  \xhookrightarrow{?(q,p,{<}m{>})} \:  \ell'_q
	      \end{align*}
	      The $ p $ and $ q $ processes  should thus be able to perform $ !(p,q,x) $ or
	      $ !(p,q,{<}m{>}) $ (in $ p $)
	      and $ ?(q,p,x) $ or $ ?(q,p,{<}m{>}) $  (in $ q $) simultaneously.
	      Then, message passing can take place between $ p $ and $ q $. The effect of message
	      passing corresponds to the (distributed) assignment $ x \coloneqq m $ (in the case of the first
	      receiving action)
	      or to removing the value $ m $ from the channel if and only if the sent value is equal to $ m $
	      (in the case of the second   receiving action).
	\item  \emph{Asynchronous message passing}. If $ cap((p,q)) > 0 $, then process $ p $ can perform the
	      conditional transition
	      \begin{align*}
		       & \ell_p  \xhookrightarrow{!(p,q,x)} \:  \ell'_p \quad \text{or} \\
		       & \ell_p  \xhookrightarrow{!(p,q,{<}m{>})} \:  \ell'_p
	      \end{align*}
	      if and only if channel $ (p,q) $ is not full, i.e., if less than $ cap((p,q)) $
	      messages are stored in $ (p,q) $.
	      In this case, $ x $ or $ m $ are stored at the rear of the buffer $ (p,q) $.
	      Channels are thus considered as
	      first-in, first-out buffers. Accordingly, $ q $ may perform
	      \begin{align*}
		       & \ell_q  \xhookrightarrow{?(q,p,x)} \:  \ell'_q \quad   or \\
		       & \ell_q  \xhookrightarrow{?(q,p,{<}m{>})} \:  \ell'_q
	      \end{align*}
	      if and only if the buffer of $ (p,q) $ is not empty. In this case, the first element  of the
	      buffer is extracted and assigned to $ x $ in an atomic manner
	      (in the case of the first receiving action) or just extracted
	      (in the case of the second   receiving action if the sent value is equal to $ m $).
	      This is summarized in Table~\ref{tab:commAction}.

\end{itemize}

\begin{table}
	\centering
	\begin{tabular}{lll}
		\toprule
		                   & executable if\dots                       & effect                                                       \\
		\midrule
		$ !(p,q,x) $       & $ (p,q) $ is not full                    & $  Enqueue((p,q),x) $                                        \\
		\midrule
		$ !(p,q,{<}m{>}) $ & $ (p,q) $ is not full                    & $  Enqueue((p,q),m) $                                        \\
		\midrule
		$ ?(q,p,x) $       & $ (p,q) $ is not empty                   & $\langle x \coloneqq Front((p,q)) ; Dequeue((p,q)) \rangle $ \\
		\midrule
		$ ?(q,p,{<}m{>}) $ & \makecell{$ (p,q) $ is not empty and the                                                                \\ sent value is equal to $ m $}  & $Dequeue((p,q)) $ \\
		\bottomrule
	\end{tabular}
	\caption{\label{tab:commAction}  Enabledness and effect of communication actions if $ cap((p,q)) > 0 $.}
\end{table}

\section{\texorpdfstring{Semantics of \glspl{tpcs} as \glspl{tpts}}{Semantics of TPCSs as TPTSs}}

\subsection{Transition System Semantics of Program Graphs}%
\label{ssec:pg_from_ts}

Intuitively, each program graph can be interpreted as a transition
system. The states of the transition system consist of pairs of
locations and variable evaluations of the program graph. Clearly, if
the domains of the variables are not finite, the resulting transition
system will have an infinite number of states. On the other hand, if
the domains of all the variables are finite, the resulting transition
system will have a finite set of states --- albeit possibly much
larger than the set of locations in the program graph. The set $AP$ of
propositions in the transition system is comprised of locations and
Boolean conditions for the
variables. In this way, whenever $l \xhookrightarrow{g:\alpha} l'$ is
a conditional transition in the program graph and the guard $g$ holds
in the current evaluation $\eta$, then there is a transition from
state $\langle l, \eta \rangle$ to state $\langle l', \Effect(\alpha,
	\eta)\rangle$. Formally the transition system $\TS(\PG)$ of a program
graph
\[
	\PG = \PGdef
\]
over the set of variables $\Var$ is the tuple $\TS(PG)=\TSdef$ where:
\begin{itemize}
	\item $S = \Loc \times \Eval(\Var)$, i.e., the same location in the program graph corresponds to several states in the transition system, each one characterized by a unique evaluation of the variables.
	\item $\longto \subseteq S \times \Act \times S$ is defined by the
	      following rule:
	      \[
		      \frac{l \xhookrightarrow{g:\alpha} l' \;\; \wedge \;\; \eta \models g}
		      {\langle l, \eta \rangle \xrightarrow{\alpha} \langle l',
			      \Effect(\eta, \alpha)\rangle}
	      \]
	      where $l,l' \in \Loc$, $g \in \Cond(Var)$, $\alpha \in \Act$ and $\eta
		      \in \Eval(\Var)$; intuitively, if the program graph features a
	      transition on action $\alpha$ from $l$ to $l'$ subject to the guard
	      $g$, the underlying transition system will have a transition on action
	      $\alpha$ from state $\langle l, \eta \rangle$ to state $\langle l',
		      \eta' \rangle$ as long as $\eta \models g$ and $\eta' = \Effect(\eta,
		      \alpha)$.
	\item $I = \{\langle l, \eta \rangle \mid l \in \Loc_0,\:\eta
		      \models g_0\}$;
	\item $\AP = \Loc \cup \Cond(\Var)$.
	\item $L(\langle l,\eta \rangle) = \{l\} \cup \{g \in Cond(Var) \mid \eta \models
		      g \}$.
\end{itemize}

Also the semantics of probabilistic and timed extensions of program
graphs can be defined in terms of transition systems. The semantics of
a probabilistic program graph $\PPG = \PPGdef$ over the set of
variables $\Var$ is a probabilistic transition system $\PTS(\PPG) =
	\PTSdef$ where:
\begin{itemize}
	\item $S = \Loc \times \Eval(\Var)$
	\item $p : S \times \Act \times S \to [0,1]$ is defined as
	      \[
		      p(\langle l, \eta \rangle, \alpha, \langle l', \eta' \rangle) = \left\{
		      \begin{array}{ll}
			      \Effect_p(\eta, \alpha, \eta') & \mbox{if } l
			      \xhookrightarrow{g:\alpha} l' \mbox{ and } \eta \models g \\
			      0                              & \mbox{otherwise}
		      \end{array}
		      \right.
	      \]
	      where $l,l' \in \Loc$, $g \in \Cond(\Var)$, $\alpha \in \Act$ and
	      $\eta \in \Eval(\Var)$;
	      This means that the probability of making a transition from state
	      $\langle l, \eta \rangle$ to $\langle l', \eta' \rangle$ is exactly
	      $\Effect_p(\eta, \alpha, \eta')$ as long as location $l$ satisfies
	      the guard to make a transition on action $\alpha$ to location $l$
	      and it is zero otherwise. Since for all variables $x \in \Var$, for
	      all evaluations $\eta(x) \in \Eval(x)$ and for all actions $\alpha
		      \in \Act$ we have
	      \[
		      \sum_{\eta'(x) \in \Eval(x)} \Effect_p(\eta(x), \alpha, \eta'(x)) = 1
	      \]
	      then it is also the case that, for all states $\langle l, \eta \rangle$
	      and for all actions $\alpha \in \Act$ we have
	      \[
		      \sum_{\langle l', \eta' \rangle} p(\langle l, \eta \rangle,
		      \alpha, \langle l', \eta' \rangle) = 1
	      \]
	\item $I = \{\langle l, \eta \rangle \mid l \in \Loc_0,\:\eta
		      \models g_0\}$
	\item $\AP = \Loc \cup \Cond(\Var)$
	\item $L(\langle l,\eta \rangle) = \{l\} \cup \{g \in \Cond(\Var) \mid \eta \models
		      g \}$
\end{itemize}

The semantics $\TTS(\TPG)$ of a timed program
graph
\[
	\TPG = \TPGdef
\]
over the set of variables $\Var$ and the set of clocks $\Clocks$ is
the tuple $\TTS(\TPG)=\TTSdef$ where:
\begin{itemize}
	\item $S = \Loc \times \Eval(\Var)$
	\item $X = \Clocks$
	\item $\Phi(X) = \Cond(\Clocks)$
	\item $\longto \subseteq S \times \Act \times 2^X \times \Phi(X) \times S$
	      is defined by the following rule:
	      \[
		      \frac{l \xhookrightarrow{g,\phi:\alpha} l' \;\; \wedge \;\; \eta
			      \models g \; \; \wedge \; \; \delta \models \phi}
		      {\langle l, \eta \rangle \xrightarrow{\alpha,Y,\phi} \langle l',
			      \Effect(\eta, \alpha)\rangle}
	      \]
	\item $I = \{\langle l, \eta \rangle \mid l \in \Loc_0,\:\eta
		      \models g_0\}$
	\item $\AP = \Loc \cup \Cond(\Var)$
	\item $L(\langle l,\eta \rangle) = \{l\} \cup \{g \in \Cond(\Var) \mid \eta \models
		      g \}$
\end{itemize}

\subsection{Transition System Semantics of Channel Systems}%
\label{ssec:pg_from_ts_comp}

Let $\varepsilon  $ denotes an empty channel, $  \ell_i $ indicates the current location of component $ \PG_i $,
$ \eta \in \Eval(\Var) $ is an evaluation of the variables, $ \xi $ is a channel evaluation,
$ \Eval(\Chan) $ denotes the set of all channel evaluations, $ len(\cdot) $ denotes the length of a sequence,
$\CS = [\PG_1 | \dots | \PG_n] $ be a channel system over $ (\Chan, \Var) $ with
\begin{align*}
	\PG_i = (\Loc_i, \Act_i, \Effect_i, \hookrightarrow_i  , \Loc_{0,i}, g_{0,i}) ,\quad\text{for } 0 < i \leqslant n.
\end{align*}

The transition system of $ \CS $, denoted $ \TS(\CS) $, is the tuple $ (S, \Act, \rightarrow, I, \AP, L) $ where:
\begin{itemize}
	\item $ S = (\Loc_1 \times \cdots \times \Loc_n) \times \Eval(\Var) \times \Eval(\Chan) $ is a set of states,
	\item  $ \Act = \biguplus_{0 < i \leqslant n}
		      \Act_i\:\uplus\:\{\tau\}$ is a set of actions ($ \{\tau\} $
	      represents a set of internal actions),
	\item  $\rightarrow$ is a transition relation defined by the rules of Figure~\ref{fig:channelRules},
	\item $ I =  \left\{ \langle \ell_1, \ldots, \ell_n,\eta,\xi_0 \rangle \: | \: \forall 0 < i \leqslant n.
		      (\ell_i \in \Loc_{0,i} \wedge \eta \vDash g_{0,i}) \right\}$ is a set of initial states,
	\item $ \AP = \biguplus_{0 < i \leqslant n} \Loc_i\:\uplus\:\ \Cond(\Var)$ is a set of atomic propositions, and
	\item $ L(\langle \ell_1, \ldots, \ell_n,\eta,\xi \rangle ) = \{\ell_1, \ldots, \ell_n\} \: \bigcup \:
		      \{g \in \Cond(\Var) \; | \; \eta \vDash g \} $ is a labeling function.
\end{itemize}

\begin{figure}[t!]
	\fbox{%
		\begin{minipage}{\textwidth}
			{\begin{itemize}
					\item interleaving for $ \alpha \in Act_i $:
					      \begin{align*}
						      \frac{\ell_i  \xhookrightarrow{g:\alpha} \:  \ell'_i \quad \wedge \quad \eta \vDash g }
						      {\langle \ell_1, \ldots,  \ell_i, \ldots, \ell_n,\eta,\xi \rangle \xrightarrow{\alpha}
							      \langle \ell_1, \ldots,  \ell'_i, \ldots, \ell_n,\eta',\xi \rangle}
					      \end{align*}
					      where $ \eta' = \Effect(\alpha,\eta) $
					\item asynchronous message passing for $ (p,q) \in Chan, cap((p,q))=1,
						      len(\xi(p,q))=1, \quad p,q \in Proc, p \neq q $
					      \begin{itemize}
						      \item receive a value $ v \in \dom((p,q)) $ along channel $ (p,q) $ and assign it to\\
						            variable $ x \in Var $  with  $ \dom(x) \supseteq \dom((p,q)) $:
						            \begin{align*}
							            \frac{\ell_i  \xhookrightarrow{g:?(q,p,x) } \:  \ell'_i \quad \wedge \quad \eta \vDash g
								            \quad \wedge \quad \xi((p,q)) = v}
							            {\langle \ell_1, \ldots,  \ell_i, \ldots, \ell_n,\eta,\xi \rangle \xrightarrow{\tau}
								            \langle \ell_1, \ldots,  \ell'_i, \ldots, \ell_n,\eta',\xi' \rangle}
						            \end{align*}
						            where $ \eta' = \eta[x \coloneqq v] $ and $ \xi' = \xi[(p,q) \coloneqq \varepsilon] $
						      \item receive a specific value  $ m \in \dom((p,q)) $ along channel $ (p,q) $
						            \begin{align*}
							            \frac{\ell_i  \xhookrightarrow{g:?(q,p,{<}m{>}) } \:  \ell'_i \quad \wedge \quad \eta \vDash g
							            \quad \wedge \quad \xi((p,q)) = m}
							            {\langle \ell_1, \ldots,  \ell_i, \ldots, \ell_n,\eta,\xi \rangle \xrightarrow{\tau}
							            \langle \ell_1, \ldots,  \ell'_i, \ldots, \ell_n,\eta,\xi' \rangle}
						            \end{align*}
						            where  $ \xi' = \xi[(p,q) \coloneqq \varepsilon] $
						      \item transmit a variable  $ x \in Var $  with  $ \dom(x) \supseteq \dom((p,q)) $
						            over channel $ (p,q)  $:
						            \begin{align*}
							            \frac{\ell_i  \xhookrightarrow{g:!(p,q,x) } \:  \ell'_i \quad \wedge \quad \eta \vDash g
								            \quad \wedge \quad \xi((p,q)) = \varepsilon}
							            {\langle \ell_1, \ldots,  \ell_i, \ldots, \ell_n,\eta,\xi \rangle \xrightarrow{\tau}
								            \langle \ell_1, \ldots,  \ell'_i, \ldots, \ell_n,\eta,\xi' \rangle}
						            \end{align*}
						            where  $ \xi' = \xi[(p,q) \coloneqq x] $
						      \item transmit a specific value $ m \in \dom((p,q)) $ over channel $ (p,q)  $:
						            \begin{align*}
							            \frac{\ell_i  \xhookrightarrow{g:!(p,q,{<}m{>}) } \:  \ell'_i \quad \wedge \quad \eta \vDash g
							            \quad \wedge \quad \xi((p,q)) = \varepsilon}
							            {\langle \ell_1, \ldots,  \ell_i, \ldots, \ell_n,\eta,\xi \rangle \xrightarrow{\tau}
							            \langle \ell_1, \ldots,  \ell'_i, \ldots, \ell_n,\eta,\xi' \rangle}
						            \end{align*}
						            where  $ \xi' = \xi[(p,q) \coloneqq m] $
					      \end{itemize}
					\item synchronous message passing over $ (p,q) \in Chan, cap((p,q))=0$:
					      \begin{itemize}
						      \item a value $ v \in \dom((p,q)) $  and assign it to
						            variable $ x \in Var $  with  $ \dom(x) \supseteq \dom((p,q)) $:
						            \begin{align*}
							            \frac{\ell_q  \xhookrightarrow{g1:?(q,p,x) } \:  \ell'_q \quad \wedge \quad \eta \vDash g_1
								            \quad \wedge \quad \eta \vDash g_2 \quad \wedge \quad \ell_p  \xhookrightarrow{g2:!(p,q,v) }
								            \:  \ell'_p
								            \quad \wedge \quad p \neq q}
							            {\langle \ell_1, \ldots,  \ell_q, \ldots,  \ell_p, \ldots, \ell_n,\eta,\xi \rangle
								            \xrightarrow{\tau}
								            \langle \ell_1, \ldots,  \ell'_q, \ldots,  \ell'_p,\ldots, \ell_n,\eta',\xi \rangle}
						            \end{align*}
						            where  $ \eta' = \eta[x \coloneqq v] $
						      \item a specific value $ m \in \dom((p,q)) $:
						            \begin{align*}
							            \frac{\ell_q  \xhookrightarrow{g1:?(q,p,{<}m{>}) } \:  \ell'_q \quad \wedge \quad \eta \vDash g_1
							            \quad \wedge \quad \eta \vDash g_2 \quad \wedge \quad \ell_p
							            \xhookrightarrow{g2:!(p,q,{<}m{>}) }
							            \:  \ell'_p
							            \quad \wedge \quad p \neq q}
							            {\langle \ell_1, \ldots,  \ell_q, \ldots,  \ell_p, \ldots, \ell_n,\eta,\xi \rangle
							            \xrightarrow{\tau}
							            \langle \ell_1, \ldots,  \ell'_q, \ldots,  \ell'_p,\ldots, \ell_n,\eta,\xi \rangle}
						            \end{align*}
					      \end{itemize}
				\end{itemize}}
		\end{minipage}}
	\caption{\label{fig:channelRules}  Rules for the transition relation of a channel system. }
\end{figure}

\section{State Charts}\label{sec:sc}

% \Glspl{sc}\dots
\Gls{scxml}\footnote{\url{https://www.w3.org/TR/scxml/}} is ``a general-purpose event-based state machine language'' with an informal semantics specification.
Multiple \gls{scxml} machines execute asynchronously and can send events (with data parameters) to each other.
Moreover, \gls{scxml} automata can refer to a datamodel for scripting and expressions.

\subsection{Subset of \Glsfmttext{scxml}}

We assume an ECMAScript datamodel, which is the same used by the official specification for its examples.
Moreover, we consider a restricted version of the \gls{scxml} language
that enforces the following constraints:
\begin{itemize}
	\item No hierarchical states.
	\item No \texttt{<parallel>} states. As with hierarchical states,
	      definining a precise semantics for parallel states is difficult.
	\item No \texttt{<final>} states. Since we admit no final
	      states, also the element \texttt{<donedata>} is not required, as it is
	      used to pass along data once a final state is reached.
	\item No \texttt{<history>} pseudo-states. Similarly to hierarchical
	      and parallel states, this construct might simplify the specification
	      in some cases, but to the best of our knowledge is not strictly
	      required, meaning that it is always possible to provide a
	      specification that maintains the required functionality without
	      requiring history pseudo-states.
	\item No \texttt{<script>} executable elements.
	\item No \texttt{<cancel>} executable elements. This element is used
	      to cancel a delayed event that was sent, but this is a ``best effort''
	      directive sent to the \gls{scxml} processor, so its result is not well
	      defined. In our setting, we assume that all the events that are sent
	      are processed.
	\item No \texttt{<invoke>} elements. These elements are used to create
	      instances of external services but no such provision is
	      needed in our use of \gls{scxml}.\@
	      For the same reason, we ban also \texttt{<content>} data.
	\item Finally, we do not consider any system variable other
	      than \texttt{\_event} and its properties (in the ECMAScript sense)
	      \texttt{\_event.data} and \texttt{\_event.origin}.
\end{itemize}

\subsection{Time and Probabilities}

Time is only referred to in the \gls{scxml} specification with the \texttt{delay} attribute of the \texttt{<send>} element.
We thus allow specifying a delay to the sending of an event in terms of time-units,
while all other steps of the execution are assumed to be instantaneous.

Probabilities are never referred to in the \gls{scxml} specification.
Thus, to allow for probabilistic behaviour without violating the specification,
we allow for use of a \texttt{Math.random()} function in the datamodel,
returning a random float value between 0 and 1.

\section{Semantics of SC as TPCS}

We now take a set of \gls{scxml} automata and build a \gls{tpcs}.
Since \gls{scxml} does not have a formal semantics,
no formal statement can be made about this translation,
but the construction is designed to carefully implement the specification,
thus effectively providing a formal semantics to \gls{scxml} in terms of \glspl{tpcs}.

To begin with,
every \gls{scxml} automaton \(A\) is assigned a unique integer \(\id_A\).
Similarly, every event \(e\) is assigned a unique integer \(\id_e\),
and it is taken note of all of its possible source and target automata.
If the event has parameters,
these are given a fixed order to be packed as a tuple.

Every \gls{scxml} automaton \(A\) gives rise to a \gls{pg} \(\PG_A\),
associated with a channel \(\qint_A\) implementing the automaton's internal queue and carrying events' ids,
and another channel implementing the automaton's external queue \(\qext_A\) and carrying pairs of integers representing the event's id and the origin automaton's id.
Moreover, each incoming event \(e\) carrying parameters is associated to a further channel \emph{per sending origin automaton} \(A'\) to deliver data, \(c_{e, A'}\).
Note that parameters cannot be passed together with their event on \(\qext_A\),
as parameters can have any type but channels' messages are homogeneous.
Note that different origin automata sending the same event to the same target automaton cannot send data on the same channel because there would be no way to guarantee that messages are enqueued in the same order that the associated events are enqueued in the external queue \(\qext_A\).
The \gls{pg} is also given:
\begin{itemize}
	\item an integer variable \(\event\) representing the id of the \texttt{<event>} currently being processed;
	\item an integer variable \(\origin\) representing the id of the origin automaton of the event currently being processed;
	\item a variable \(p^e_i\) of matching type for each parameter of each event \(e\) that the automaton can receive;
	\item a variable \(x_l\) of matching type for each location \(l\) of the automaton's datamodel.
\end{itemize}

\subsection{States and transitions}

Each \texttt{<state>} \(s\) of the \gls{scxml} automaton is associated to a location \(p_s\) of the \gls{pg}.
If the \texttt{<onentry>} element is present,
its executable content is executed immediately
via a sequence of transitions terminating in a new location \(p'_s\).

Starting from \(p'_s\), for each eventless \texttt{<transition>} \(t\) on this state,
let \(p_t\) be the current location.
Consider \(t\)'s guard expression, convert it into a \gls{tpcs} expression \(g\),
and use it and its negations as guards for two transitions,
moving to new locations representing that \(t\) is, respectively,
enabled (\(\hat{p}_t\)) and not enabled (\(p_{t'}\), where \(t'\) is the next event).
\[
	\hat{p}_{t} \xhookleftarrow{g} p_t \xhookrightarrow{\lnot g} p_{t'}
\]

The location \(\hat{p}_{t}\) representing that the \texttt{<transition>} \(t\) is enabled
is the beginning of a chain of transitions implementing, in order:
\begin{itemize}
	\item the executable content inside the state's \texttt{<onexit>} element, if present;
	\item the executable content inside the \texttt{<transition>} element, if present;
\end{itemize}
and which ultimately reaches the location \(p_{s'}\) representing the target \texttt{<state>} \(s'\) of the \texttt{<transition>},
given by its \texttt{target} attribute.

The location representing that the \texttt{<transition>} \(t\) is not enabled
is also the location testing for the next eventless \texttt{<transition>} \(t'\),
if any, or to a new location \(p\) representing the processing of events,
which takes a transition with the effect of reading the next event from the automaton's internal queue, if any,
or another transition with the effect of reading the next pair of event and origin automaton's id from the automaton's external queue,
plus the associated parameters' data.
This is built as
\[
	p_t \xhookleftarrow{?(\event, \qint_A)} p
	\xhookrightarrow{\qint_A \colon \emptyset} p'
	\xhookrightarrow{?((\event, \origin), \qext_A)} p''
	\xhookrightarrow{\event = \id_e \land \origin = \id_{A'} \colon ?((p^e_1, \dots, p^e_n), c_{e, A'})} p_t
\]
where \(e\) ranges over the events that \(A\) may receive,
and \(A'\) over all of possible automaton sending \(e\).
Note that this requires being able to test a channel for emptiness.
Both of these transitions move to the location \(p_t\) representing the first eventful \texttt{<transition>} \(t\) of the automaton's \texttt{<state>}.

Then, the construction for eventful transitions proceeds analogously to the construction for eventless \texttt{<transition>}s,
with the addition that the guards also have to check that the \texttt{<transition>}'s event corresponds to the one currently being processed,
which amounts to testing for equality the variable representing the id of the event currently being processed and the id itself,
which is a constant integer value.
\[
	\hat{p}_{t_{e}} \xhookleftarrow{\event = \id_e \land g} p_{t_e} \xhookrightarrow{\lnot (\event = \id_e \land g)} p_{t_{e'}}
\]
When there are no eventful \texttt{<transition>}s left,
the ``not enabled'' branch leads back to the event-reading location \(p\).
Note that,
in case no \texttt{<transition>} is enabled and no event is enqueued,
the automaton keeps waiting for the arrival of a new external event.
Finally, note that transitions are processed in the order prescribed by the \gls{scxml} specification.

\subsection{Executable content}

Executable content is processed sequentially,
expressing each step via one or more transitions from a starting location to a target location.
We assume that every expression in the datamodel can be translated into a corresponding expression of the \gls{pg}
(thus limiting the expressivity of the datamodel)
where, in particular, each variable of the datamodel is associated to a variable of the \gls{pg}.

The executable elements are processed as follows:
\begin{itemize}
	\item \texttt{<assign>} is translated as an effect of the transition,
	      where the variable associated to the \texttt{location} is assigned the expression
	      that represents \texttt{expr}.
	\item \texttt{<if>}, \texttt{<elseif>} and \texttt{<else>} are translated using guards to enforce the choice of different branches;
	      each branch executes the associated executable content,
	      before converging to a common final state.
	\item \texttt{<raise>} is translated as the sending the id associated to the event as a message to the automaton's internal events' queue.
	\item \texttt{<send>} is translated as the sending of a message to the \texttt{target}'s external events' queue,
	      comprising the ids of the event and of the sender automaton,
	      followed by the sending of another message containing the event's parameters, if any.
	      If the \texttt{<send>} element has a dynamic target given by a \texttt{targetexpr},
	      the sending of the message is repeated in parallel for each possible receiver,
	      and is guarded by the condition that the receiver is the expected target.
	      Moreover, if present, the \texttt{delay} attribute is translated as a pair of time guard and invariant that enforces that the transition is performed exactly after the given number of time-steps.
\end{itemize}

% \section{JANI automata and CS for JANI (CS4J)}

% content\dots

% \section{Semantics of JANI as CS4J}

% content\dots

%%%%%%%%%%%%%%%%%%%%%%%%%
\section{\texorpdfstring{\glsfmtshort{scan}}{SCAN}}%
\label{sub:scan}
%%%%%%%%%%%%%%%%%%%%%%%%%

\begin{figure}[tb]
	\begin{center}
		{\ttfamily \input{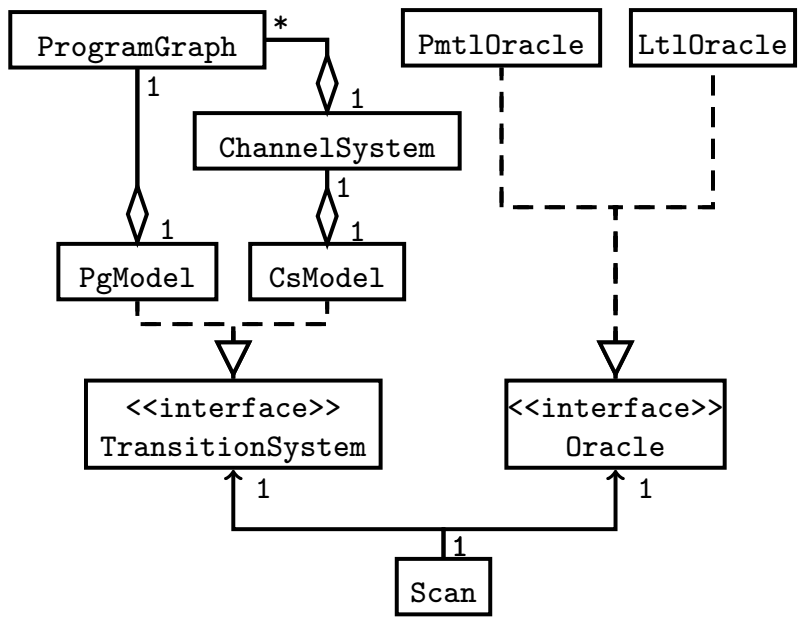}}
	\end{center}
	\caption{\label{fig:scan_class}Simplified class diagram for \gls{scan}.}
\end{figure}

\href{https://convince-project.github.io/scan/}{\Gls{scan}}~\cite{scan_home} is an in-development
statistical model checker specifically designed to verify large concurrent models of
cyber-physical systems.
It is open-source and hosted on a
\href{https://github.com/convince-project/scan}{GitHub repository}~\cite{scan_repo}.
It is cross-platform but it currently has to be installed from sources.
Its architecture comprises the following packages:
\begin{description}
	\item[\texttt{scan\_core}] implements classes for the
	      modeling formalisms and verification algorithms, abstracting away the
	      input model and property description language.  The package
	      implements a modular architecture based on interfaces to abstract
	      over different kinds of models, as shown in
	      \autoref{fig:scan_class}. In particular, \texttt{TransitionSystem}
	      is an interface abstracting a generic model of computation, and an
	      \texttt{Oracle} is an interface abstracting a generic oracle over an
	      execution trace for some temporal logic.  Together, they provide a
	      general framework to execute models and perform verification of
	      properties in the relevant logic.
	\item[\texttt{scan\_\textit{lang}}] language packages support the
	      corresponding \textit{lang} model description languages while
	      relying on \texttt{scan\_core} for modeling and verification
	      functionalities. At the moment, we implement
		      {\ttfamily scan\_scxml} to support the \gls{scxml}
	      standard as described in Section~\ref{sec:sc}
	      and {\ttfamily scan\_jani} to provide (limited) support to the
	      JANI format.
	      A package {\ttfamily scan\_promela} to support the Promela language~\cite{holzmann1997model} is under development but not yet usable.
	\item[\texttt{scan}] implements a command line user interface.
\end{description}

\Gls{scan}'s internal representation of a system implements a \gls{tpcs} as defined in \autoref{sec:tpcs}.
On top of these representations, \gls{scan} implements
\texttt{CsModel}, i.e., the model of computation for
channel systems, whereby properties are restricted to
predicate over the messages being exchanged through the channels, and
\texttt{PgModel}, i.e., a model of computation for
program graphs, whereby properties can predicate on all
state variables.
\gls{scan} also implements two oracles for checking properties:
\texttt{PmtlOracle}, an oracle for pMTL properties over state-event
traces~\cite{sarjoughian2015superdense} and discrete (but
dense~\cite{chaki2004state-event}) time; and \texttt{LtlOracle}, a
simpler oracle for (a subclass of)  LTL properties.

%Algorithm
Combining a transition system and an oracle into a \texttt{Scan}
object, it is possible to apply abstracted statistical model checking
verification algorithms.
Specifically, \gls{scan} uses the New Adaptive Sampling
Method~\cite{adaptive_chen_xu} to estimate a success rate with the
given confidence and precision. Non-determinism is currently modeled
as a uniform distribution across alternatives, though more advanced
algorithms are planned. Samples are, by default, generated through
multi-threaded parallel computation, and execution is stopped as soon
as all of the provided properties are verified to hold (such as an
``eventually'' property holding at one point in time), or one is
verified to have been violated (such as an ``always'' property not
holding at one point in time), or the model's execution is completed
(at which point all temporal properties are either true or  false).

Model specifications are parsed by the appropriate language-supporting library,
then converted into some \texttt{TransitionSystem} and \texttt{Oracle}
to be verified, while retaining model-specific data to be able to
reconstruct meaningful execution traces.
%% \begin{figure}[htb]
%% 	% \begin{adjustbox}{width=\textwidth}
%% 	{\ttfamily
%% 		\input{Diagrams/process}}
%% 	% \end{adjustbox}
%% 	\caption{\label{fig:scan_process}Simplified process diagram for \SCAN.}
%% \end{figure}
%% \MKtodo{Do we remove this picture? Fig. 3}
A module \texttt{scan\_scxml} provides support for (a subset of) the \gls{scxml}
modeling language with the ECMAScript data model.
It implements a parser and a model builder that turns a set of \gls{scxml}
specification into a channel system, one program graph per state machine.
Together with pMTL properties specified in a suitable XML format,
this results into a \texttt{CsModel} transition system and a
\texttt{PmtlOracle} oracle. Other than statistical property
verification, \gls{scan} can output traces that list the exchange of \gls{scxml}
events (and their parameters) between FSMs, so that the internal
representation of the model is transparent to the user.
A module \texttt{scan\_jani} provides experimental support for the
JANI modeling language. It implements a parser and a model builder
that turns a JANI model specification into a single program graph,
where each JANI FSM is executed as a synchronized process.
Only simple properties of the form $p \Until q$, where $p,q$ are atomic
propositions, are supported at the moment. This results into a
\texttt{PgModel} transition system and an \texttt{LtlOracle}.
Other than statistical property verification, \gls{scan} can output traces
that list the states of the FSMs variables and the actions being
applied.

\gls{scan}'s design choices are motivated by the CONVINCE use cases: large
robotic systems made of many asynchronous processes.
The \gls{scxml} modeling language has been chosen for its expressivity and
because it can be specified via a graphical notation,
which makes it suitable to be used by robotics developers with no
training in formal methods.
The restriction of verifying only messages passing through channels is
motivated by the observation that the internal state of robotic
systems' components is not known directly but only through the
monitoring infrastructure fed by messages sent via some middleware, such
as ROS 2.  The representation of such a system requires a model of
computation with asynchronous process execution and message passing,
hence the choice of channel systems.
To align verification with monitoring, we chose to use the same
temporal logic used by the monitoring system of our use cases,
ROSMonitoring with Reelay oracles~\cite{ferrandorosmonitoring,ulus2025onlinemonitoringmetrictemporal},
i.e., pMTL over finite traces, which is expressive enough to handle
all the requirements in CONVINCE use cases.

% \begin{credits}
% 	\subsubsection{\ackname} This work was funded by the European Union under the Horizon Europe grant 101070227 (CONVINCE).

% 	\subsubsection{\discintname}
% 	The authors have no competing interests to declare that are
% 	relevant to the content of this article.
% \end{credits}

% \bibliographystyle{splncs04}
\printbibliography{}

@book{baier2008principles,
	title = {Principles of model checking},
	author = {Baier, Christel and Katoen, Joost-Pieter},
	year = {2008},
	publisher = {MIT press},
}

@book{puterman2009markov,
	title = {Markov decision processes: discrete stochastic dynamic programming},
	author = {Puterman, Martin L},
	volume = {414},
	year = {2009},
	publisher = {John Wiley \& Sons},
}

@article{DBLP:journals/tcs/AlurD94,
	author = {Rajeev Alur and David L. Dill},
	title = {A Theory of Timed Automata},
	journal = {Theor. Comput. Sci.},
	volume = {126},
	number = {2},
	pages = {183--235},
	year = {1994},
}

@article{Fearnhead_2018,
	year = 2018,
	month = {aug},
	publisher = {Institute of Mathematical Statistics},
	volume = {33},
	number = {3},
	author = {Paul Fearnhead and Joris Bierkens and Murray Pollock and Gareth O.
	          Roberts},
	title = {Piecewise Deterministic Markov Processes for Continuous-Time Monte
	         Carlo},
	journal = {Statistical Science},
}

@inproceedings{ferrandorosmonitoring,
	author = "Ferrando, Angelo and Cardoso, Rafael C. and Fisher, Michael and
	          Ancona, Davide and Franceschini, Luca and Mascardi, Viviana",
	editor = "Mohammad, Abdelkhalick and Dong, Xin and Russo, Matteo",
	title = "ROSMonitoring: A Runtime Verification Framework for ROS",
	booktitle = "Towards Autonomous Robotic Systems",
	year = "2020",
	publisher = "Springer International Publishing",
	address = "Cham",
	pages = "387--399",
}

@misc{ulus2025onlinemonitoringmetrictemporal,
	title = {Online Monitoring of Metric Temporal Logic using Sequential
	         Networks},
	author = {Dogan Ulus},
	year = {2025},
	eprint = {1901.00175},
	archivePrefix = {arXiv},
	primaryClass = {cs.LO},
	url = {https://arxiv.org/abs/1901.00175},
}

@inproceedings{chaki2004state-event,
	author = "Chaki, Sagar and Clarke, Edmund M. and Ouaknine, Jo{\"e}l and
	          Sharygina, Natasha and Sinha, Nishant",
	editor = "Boiten, Eerke A. and Derrick, John and Smith, Graeme",
	title = "State/Event-Based Software Model Checking",
	booktitle = "Integrated Formal Methods",
	year = "2004",
	publisher = "Springer Berlin Heidelberg",
	address = "Berlin, Heidelberg",
	pages = "128--147",
	isbn = "978-3-540-24756-2",
}

@inproceedings{sarjoughian2015superdense,
	author = {Sarjoughian, Hessam S. and Sundaramoorthi, Savitha},
	title = {Superdense time trajectories for DEVS simulation models},
	year = {2015},
	isbn = {9781510801059},
	publisher = {Society for Computer Simulation International},
	address = {San Diego, CA, USA},
	booktitle = {Proceedings of the Symposium on Theory of Modeling \&
	             Simulation: DEVS Integrative M\&S Symposium},
	pages = {249–256},
	numpages = {8},
	location = {Alexandria, Virginia},
	series = {DEVS '15},
}

@inproceedings{adaptive_chen_xu,
	title = {A new adaptive sampling method for scalable learning},
	author = {Chen, Jianhua and Xu, Jian},
	booktitle = {Proceedings of the International Conference on Information and
	             Knowledge Engineering (IKE)},
	pages = {1},
	year = {2013},
	organization = {The Steering Committee of The World Congress in Computer
	                Science, Computer~…},
}

@article{holzmann1997model,
	title = {The model checker SPIN},
	author = {Holzmann, Gerard J.},
	journal = {IEEE Transactions on software engineering},
	volume = {23},
	number = {5},
	pages = {279--295},
	year = {1997},
	publisher = {IEEE},
}

@online{scan_repo,
	editor = "{Enrico Ghiorzi}",
	title = "{SCAN}",
	url = "https://github.com/convince-project/scan",
	urldate = "2025-06-20",
}

@online{scan_home,
	editor = "{Enrico Ghiorzi}",
	title = "The {SCAN} Book",
	url = "https://convince-project.github.io/scan/",
	urldate = "2025-06-20",
}
\end{document}